\documentclass[aps,prb,amsmath,amssymb,footinbib,showpacs,twocolumn,superscriptaddress
]{revtex4-2}
\usepackage{orcidlink}
\usepackage{graphicx}
\usepackage{physics}
\usepackage{comment}
\usepackage{subcaption}
\usepackage{dcolumn}
\usepackage{bm}
 
\usepackage{blkarray}
\usepackage{amsmath,calc}
\usepackage{tikz}
\usetikzlibrary{mindmap}

\begin{document}
\preprint{APS/123-QED}

\title{Tight-Binding Energy-Phase Calculation for Topological Josephson Junction Nanowire Architecture}

\author{Adrian D. Scheppe\orcidlink{0000-0002-5566-2744}}\email{adrian.scheppe.1@au.af.edu}
\author{Michael V. Pak}\email{michael.pak.4@au.af.edu}
\affiliation{Department of Physics, Air Force Institute of Technology\\
2950 Hobson Way, Wright-Patterson AFB, Ohio 45433}
\date{\today}

\begin{abstract}
The current state of Quantum computing (QC) is extremely optimistic, and we are at a point where researchers have produced highly sophisticated quantum algorithms to address far reaching problems. However, it is equally apparent that the noisy quantum environment is a larger threat than many may realize. The noisy intermediate scale quantum (NISQ) era can be viewed as an inflection point for the enterprise of QC where decoherence could stagnate progress if left unaddressed. One tactic for handling decoherence is to address the problem from a hardware level by implementing topological materials into the design. In this work, we model several Josephson junctions that are modified by the presence of topological superconducting nanowires in between the host superconductors. Our primary result is a numerical calculation of the energy-phase relationship for topological nanowire junctions which are a key parameter of interest for the dynamics of qubit circuits. In addition to this, we report on the qualitative physical behavior of the bound states as a function of superconducting phase. These results can be used to further develop and inform the construction of more complicated systems, and it is hopeful that these types of designs could manifest as a fault tolerant qubit.
\end{abstract}
\maketitle
\section{Introduction:}
Quantum computing (QC) and quantum information (QI) have developed into an enormous worldwide enterprise that has attracted the attention from a wide array of investors, ranging from commercial and academic parties to political and military entities \cite{parker2021commercial,bova2021commercial,brandmeier2022future,srivastava2016commercial,bayerstadler2021industry,krelina2021quantum,neumann2020quantum,smith2020quantum}. Indeed, due to the many impressive results in QC algorithms, this disruptive technology is currently of high scientific, corporate, and strategic interest for the involved parties \cite{mavroeidis2018impact,cao2018potential,gyongyosi2019survey,perdomo2018opportunities,wu2021strong}. In retrospect, this is an ironic turn of events, because, when QC was first imagined, it was thought that \textit{any} two level system could realize a workable qubit. However, it was not clear for what such a thing could be used. That is, there were no algorithms. Now, the opposite is true where one can easily visualize a purpose for this technology, but the list of practically useful hardware has dwindled to only a handful of options.

At this stage, we see that there is a tremendous atmosphere of optimism for the future emenating from the intersection of physics, computer science, and mathematics and a somewhat fanatic obsession emerging from the corporate and political sectors. Physicists can afford to possess this outlook due in large part to the highly successful hardware that has been developed within the last two decades. By a large margin, the family of superconducting (SC) qubits have become the best combination of stability, flexibility, and practicality, forming the backbone of recent high caliber state-of-the-art QC and QI systems \cite{kjaergaard2020superconducting,devoret2004superconducting}. Some examples include IBM's Osprey chip with 433 qbs \cite{patra2024efficient}, Google's Sycamore chip with 53 qbs \cite{pan2022solving}, and Rigetti's Ankaa-2 chip with 84 qbs \cite{maciejewski2024improving}. There exists other types of qubits such as photonic, trapped ion, and spin, but it is clear that the SC qubits have taken center stage due most likely to their compact chip design.

\begin{figure}[b!]
    \centering
    \includegraphics[width = \columnwidth]{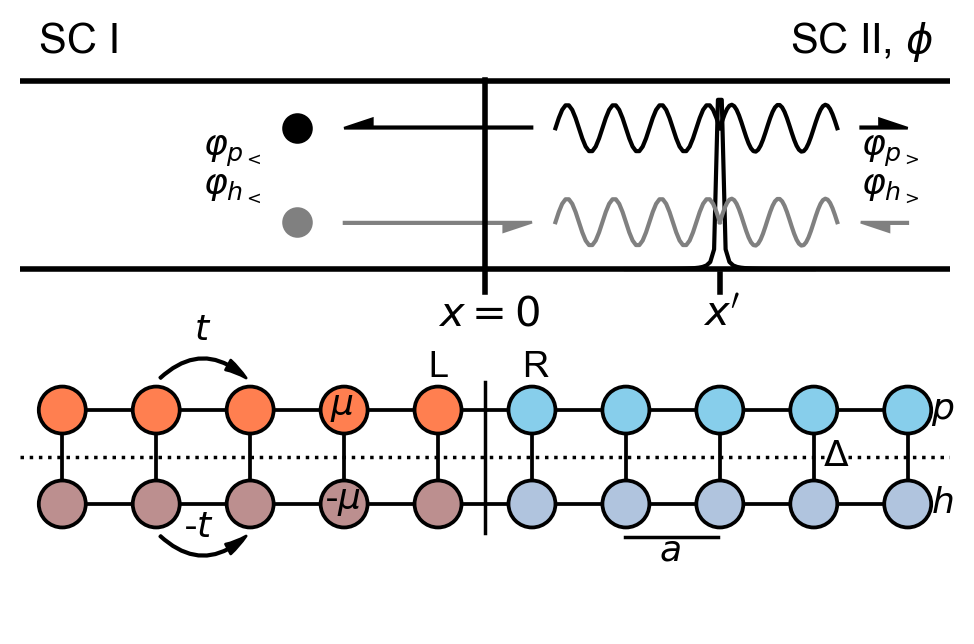}
    \caption{Normal Josephson Junction Setup. (Upper) Two almost identical 1D SC systems are placed adjacently with phase difference, $\phi$. The Green's function is made from rudimentary wavelet processes which are the result of a $\delta$-source within the system. (Lower) The tight-binding calculation we employ has two orbitals per site, lattice parameter $a$, onsite $\mu$, hopping $t$, and SC pairing potential $\Delta$. }
    \label{fig:scscSetup}
\end{figure}
Certainly, the 21st century has enjoyed large advances in SC material modeling, and annealing/ lithographic processes \cite{last2023key,sharma2022comprehensive,scheppe2023perturbing}, a highly useful tool for the creation of precise Josephson junction (JJ) architecture within microchips. Thus, engineers have a high degree of control over the physical qubit circuit design. Additionally, these qubits demonstrate fast gates and high state fidelity \cite{kjaergaard2020superconducting,krantz2019quantum}. These qualities in combination make for a highly promising option for the future of qubit technology with the main drawback being that they do not remain coherent as long as some of their counterparts. Recent benchmarks have recorded coherence times at 0.3 milliseconds \cite{place2021new} and 0.5 milliseconds \cite{wang2022towards} for the Tantulum based transmon SC qubit. Compare this to the ion trap $^{171}\text{Yt}^+$ qubit with a coherence time exceeding 10 minutes \cite{wang2017single}. Regardless of this weakness, these benchmark coherence times are increasing across the board yearly with many explaining the trend as a QC analogue to the classical ``Moore's law" \cite{ezratty2023there}. It is true that \textit{individual} qubits are performing better and better, but, as a \textit{collective}, not so much. Qualitatively, increasing the qubits within a system drastically \textit{decreases} the coherence time of the system as a whole, and each subsequent gate operation decreases it further. For example, the Osprey chip can afford to have so many qubits solely because they are far less connected compared to the grid layout used by Google and Rigetti. So, we really should only show cautious optimism for the future. We are not out of the NISQ woods yet.

One well known general stategy for qubit fault tolerance incorporates materials with nontrivial band topology into the qubit design \cite{freedman2003topological}. The motivation for this tactic stems from the fact that these systems contain physical parameters that are directly related to global properties of the system rather than local. Thus, local environmental perturbations are not enough to disturb these topological invariants. In earlier topological QC proposals, one might see the use of Abrikosov vortices on 2D topological superconductors (TSC) \cite{lahtinen2017short, scheppe2022complete} or the direct manipulation of chiral surface states \cite{lian2018topological}. While highly intriguing, these types of systems which deal with the Majorana bound states (MBS) directly do not seem at this point in time to be practically viable in the long run. This is due to a handful of engineering concerns such as a speed of vortex manipulation \cite{grimaldi2015speed} and size of system required for adequate state seperation \cite{posske2020vortex}. More concerning though, it is not clear whether or not a vortex MBS can be reliably detected \cite{castelvecchi2021evidence,frolov2021quantum} which is required for qubit readout for this style of topological QC (TQC).

A more promising scheme is to use systems comprised of TSC nanowires which modify the standard Josephson junction (JJ) \cite{schrade2018majorana,xue2013tunable,jiang2011interface,mishmash2020dephasing,knapp2018dephasing,szechenyi2020parity,karzig2021quasiparticle,schmidt2012decoherence,li2018four,karzig2017scalable}. These nanowires provide channels with innate topology to bridge the physical gap between superconductors in such a way that supercurrent is constrained and made immune to certain perturbations. These options are a more attractive bunch for two main reasons. First, by making slight topological modifications to the SC-SC junction, one is relying on the highly successful SC qubit design and all the positive attributes mentioned above. Since there is a deep understanding for these types of systems, we have the facilities and theory to probe and compare how modification affects the overall performance. Secondly, the utility of such a system does not rely on the direct detection of a MBS. Instead, the protected edge states merely facilitate robust electron transport through the junction, coupling the qubit and topological state \cite{fu2007topological}, and the readout of the qubit is the same as for standard SC qubit systems. 

One aspect of these devices that is lacking in the literature though is the dilineation of the JJ energy phase relation, $E(\phi)$. This parameter is key to deriving qubit dynamics for the SC qubits because it enters the circuit Hamiltonian which is then solved to derive the spectrum as well as perturbed to find gate operations on the computational submanifold \cite{blais2021circuit}. In all SC qubits, the specific circuit designs may become quite complicated, but, at minimum, they all must contain at least one standard JJ \cite{devoret2004superconducting,kjaergaard2020superconducting}. For the standard JJ with small barrier potential, this quantity is widely known to be $\propto -\text{cos}(\phi)$. However, for highly complicated topological architectures, this quantity and its behavior are unknown, and, for this reason, the goal of this work is to answer the question: how does the presence of 1D TSC nanowires affect the energy-phase relationship of the standard JJ?

To answer this question, we begin by outlining the general behavior for a standard 
JJ ABS. This is accomplished in Sect. \ref{sect:SCSCjunction} by a Green's function approach to analytically derive the energy-phase parameter which we use to verify a comparable tight-binding calculation. In Sect. \ref{sect:topoJuncs}, we apply this same method to a number of topological junctions to calculate the phase dependence of in-gap states.

\section{SC-SC Junction Bound States}
\label{sect:SCSCjunction}
\subsection{Continuous Model}
As a demonstration, we set out to derive an analytical form of the JJ ABS energy phase relationship. In the literature, a standard calculation of this type would be a scattering matrix formulism, where scattering coefficients are calculated by imposing boundary conditions on the total state and its derivative at the interface. However, instead of this route, we choose to proceed down different lines of reasoning by employing a Green's function approach. This is done for two main reasons. First, the logic behind assigning bound state status to poles in the scattering matrix problem stems directly from the singularities in the Green's function for an operator. Second, at present, we are not aware of any works that use this approach \textit{specifically} to calculate $E(\phi)$ for the SC-SC ABS. Though, it is used for graphene/ normal-SC interface models \cite{mcmillan1968theory,herrera2011dirac}. We believe this type of calculation to be more versatile, fruitful, and relevant for the development of SC junction engineering.

To begin, imagine two nearly identical 1D, s-wave SC chains which are out of phase with one another by $\phi$, see upper image within Fig. \ref{fig:scscSetup}. The cooper pairs within our system must be spin singlets, but other than that spin can be neglected in order to calculate parameters of interest. One can model this with a spatially dependent order parameter $\Delta(x)=\Delta_0 e^{i\phi \Theta(x)} $, where $\Delta_0$ is a constant number and $\Theta(x)$ is the Heaviside function. The Bogoliubov-de Gennes (BdeG) Hamiltonian matrix corresponding to this gap parameter is written as,
\begin{gather}
    \hat{H}=
    \begin{pmatrix}
        \text{-}\frac{\hbar^2}{2m}\frac{\partial^2}{\partial x^2}+\mu&\Delta(x)\\
        \Delta^*(x)&\frac{\hbar^2}{2m}\frac{\partial^2}{\partial x^2}-\mu
    \end{pmatrix}
    ,
\end{gather}
\begin{figure}[t!]
    \centering
    \includegraphics[width=\columnwidth]{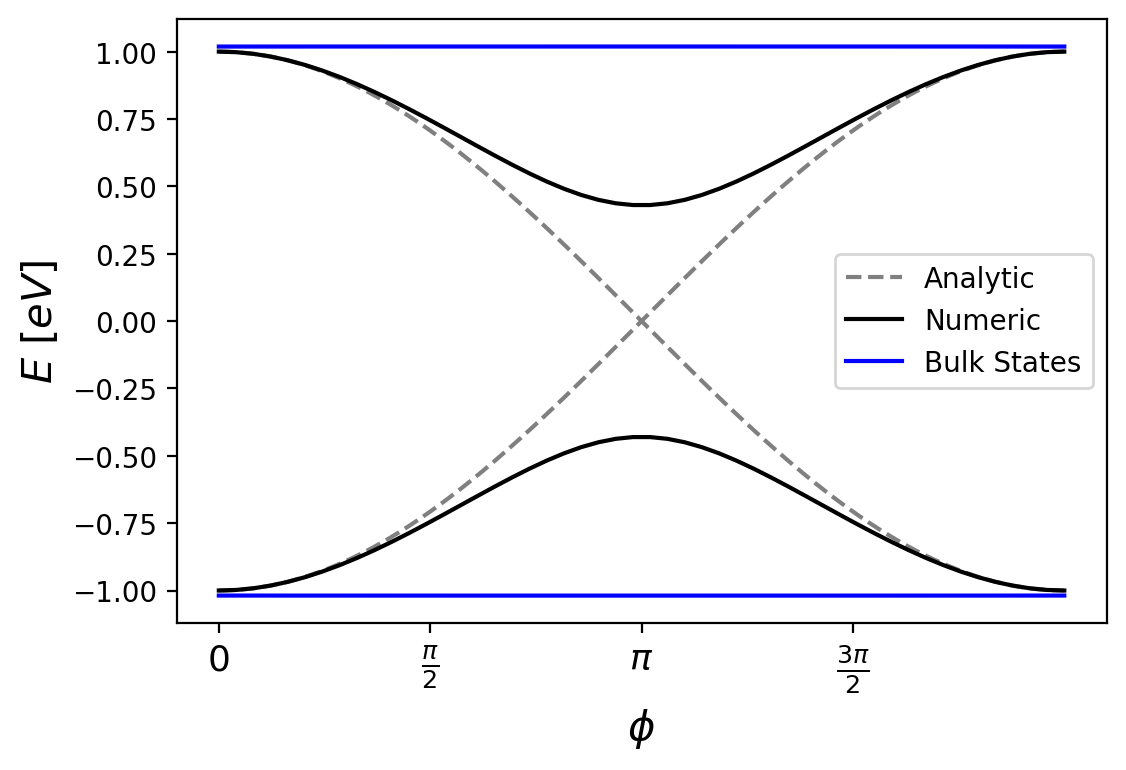}
    \caption{SC-SC Numerical Calculation of $E(\phi)$. We plot the analytic solution to the normal JJ here alongside the tight-binding calculation. The gap in the numerical calculation emerges due to the finite value of $t$. As the junction hopping parameter approaches infinity, the gap for this calculation would approach zero.}
    \label{fig:SCSC_ABS_numEn}
\end{figure}
where $\mu$ is the chemical potential. To construct the Green's function, we need the solution to the inhomogeneous differential equation, $\mathcal{L}_E(\hat{G}(x,x';E))=\delta(x-x')\mathbb{I}$, where $\mathcal{L}_E\equiv(\hat{H}-E\mathbb{I})$. This is accomplished using the tensor version of the Sturm-Louiville approach in Refs. \cite{herrera2022green, mcmillan1968theory} and outlined in Appendix \ref{app:SLGenMeth}. For this, we are required to find solutions to the matrix equation, $\hat{H}\Psi_{\lessgtr}(x)=E\Psi(x)_{\lessgtr}$, where $\Psi(x)_{\lessgtr}=(u(x),v(x))^T$ is a solution carrying left (right) boundary conditions. With asymptotic solutions, we are effectively using left and rightward travelling waves to build $\hat{G}(x,x';E)$. Assuming plane wave character in an infinite left and right system such that $\Psi(x)=e^{ikx}(u_0,v_0)^T$, one solves for $E^2=(\frac{\hbar^2}{2m}k^2_\pm-\mu)^2+\Delta_0^2$ and $k^2_\pm=k_f(1\pm\sqrt{(E^2-\Delta_0^2)/\mu^2})$ for $k_f=\sqrt{2m\mu}/\hbar$ with these quantities being identical on either side of the interface. At this point, we use the Andreev approximation which assumes $\mu\gg E,\Delta_0$ such that the wavenumber can be approximated as $k_\pm\cong k_f$. With that, the particle ($+$)/ hole ($-$)-like solutions for a \textit{right} travelling wave in regions 1 and 2 become,
\begin{equation}
\Psi_{1,2}^{p,h}(x) = e^{\pm ik_f x}\varphi^{p,h}_{1,2},\\
\end{equation}
where,
\begin{gather}
    \varphi^{p}_{1,2}=\begin{pmatrix}
        u_0\\
        v_0e^{i\phi_{1,2}}
    \end{pmatrix},\\
    \varphi^{h}_{1,2}=\begin{pmatrix}
        v_0\\
        u_0e^{i\phi_{1,2}}
    \end{pmatrix}
    \label{eq:spinor2},
\end{gather}
are the quasiparticle part of the wave function with,
\begin{align}
    u_0&=\sqrt{\frac{1}{2}(1+\sqrt{1-(\frac{\Delta}{E})^2)}},\label{eq:u}\\
    v_0&=\sqrt{\frac{1}{2}(1-\sqrt{1-(\frac{\Delta}{E})^2)}}.
    \label{eq:v}
\end{align}
Using these solutions, we construct the Green's function by the primitive left/right processes displayed in Fig. \ref{fig:scscSetup}. This is analogous to what is done Appendix \ref{app:deltaPot}, and, following the logic found there, one recognizes that the Green's function is piecewise in both $x$ and $x'$, meaning the full function will have six pieces based on the relative values of $x,x'$ with respect zero. However, we are only interested in the analytic structure of $\hat{G}$ which can be derived by analyzing the jump condition when $x=x'=0$. Thus, we construct a Green's function using leftward and rightward propagating particle/ hole-like in region 1 and 2 respectively to make the following general form,
\begin{align}
    \hat{G}(x,x'&;E) = \sum_{\alpha \beta}A_{\alpha\beta}\Psi_{1,<}^\alpha(x)\otimes \Psi_{2,>}^\beta(x'),
    \label{eq:genGreen}
\end{align}
with the various coefficients $A_{\alpha\beta}$ representing the probability of a spinor type $\alpha$ propagating from left of the barrier to right into a $\beta$ type spinor. This equation corresponds to the $x<0<x'$ piece of the total Green's  function. We can find the $x'<0<x$ piece by swapping the variables and transposing, $\hat{G}(x',x; E)^T$. These coefficients are fixed by the jump condition at $x=x'=0$ established by integrating $\mathcal{L}_E(\hat{G}(x,0;E))=\delta(x)\mathbb{I}$ with respect to $x$ from $-\epsilon$ to $\epsilon$. After taking a limit with respect to $\epsilon$, the $E$ and $\Delta(x)$ pieces of the equation go to zero, so we only need,
\begin{equation}
    -\frac{\hbar^2}{2m}\int_{-\epsilon}^{\epsilon} dx \partial^2_x\hat{G}(x,0)=\tau_z.
\end{equation}
which produces the jump condition at $x=x'=0$,
\begin{equation}
    \partial_x\hat{G}(0,x)^T|_{0}-\partial_x\hat{G}(x,0)|_{0}=\text{-}\frac{2m}{\hbar^2}\tau_z.
    \label{eq:jumpCond}
\end{equation}
Taking the derivative on the first component of the tensor product in Eq. \ref{eq:genGreen} and evaluating for $x=x'=0$, Eq. \ref{eq:jumpCond} expands to, 
\begin{widetext}
\begin{align}
    A_{pp}(\varphi^p_2\otimes\varphi^p_1+\varphi^p_1\otimes\varphi^p_2)-A_{ph}(\varphi^p_2&\otimes\varphi^h_1+\varphi^p_1\otimes\varphi^h_2)\nonumber\\&+A_{hp}(\varphi^h_2\otimes\varphi^p_1+\varphi^h_1\otimes\varphi^p_2)+A_{hh}(\varphi^h_2\otimes\varphi^h_1+\varphi^h_1\otimes\varphi^h_2)=\frac{2mi}{k_f\hbar^2}\tau_z.
\end{align}
\end{widetext}

Solving this system of equations yields the following assignments for the coefficients,
\begin{align}
    &A_{pp,hh}=\frac{im}{\hbar^2k_f}\frac{e^{-i\phi}}{u_0^2-v_0^2},\\\nonumber\\
    &A_{hp}= A_{ph}\frac{e^{i\phi}u^2_0-v^2_0}{e^{i\phi}v^2_0-u_0^2},
\end{align}
\begin{figure}[b!]
    \centering \includegraphics[width=\columnwidth]{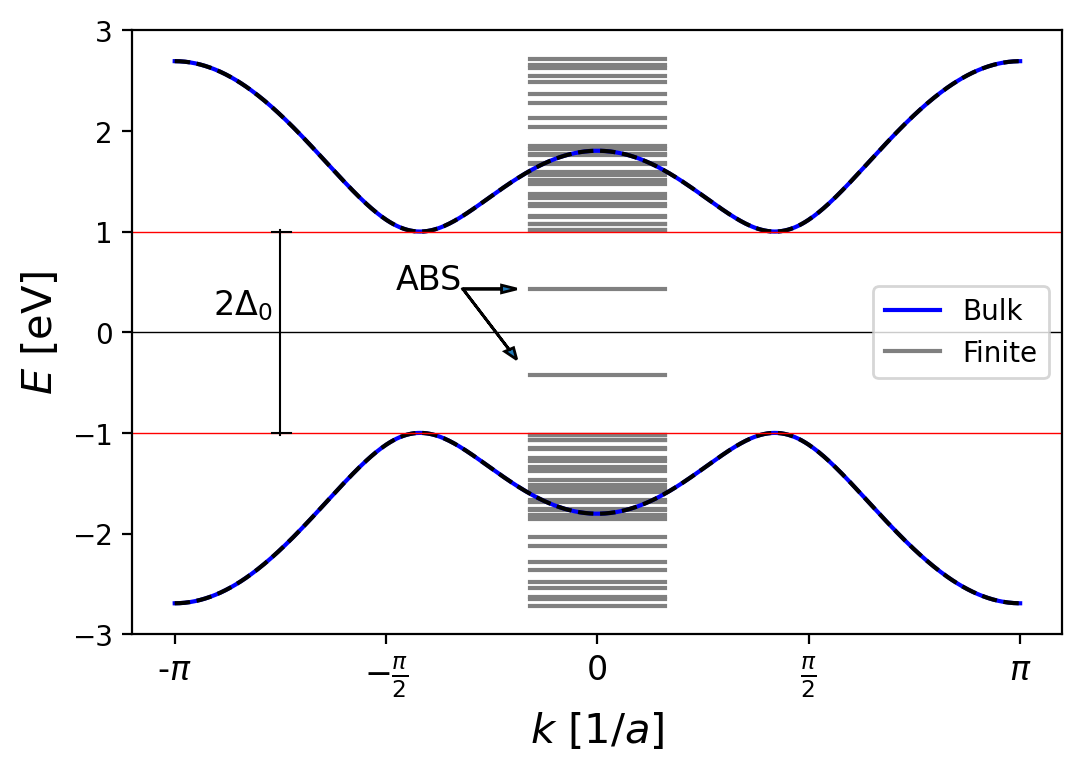}
    \caption{Bulk Dispersion vs. Finite System. In-gap states for the tight-binding system are found by comparing the finite system's eigenvalue spectrum to the bulk dispersion curves. There are two in-gap, phase sensitive states here which are ABS.}
    \label{fig:SCSC_specComp}
\end{figure}
where a pole exists whenever the following equation is satisfied,
\begin{equation}
    e^{i\phi}v^2_0-u_0^2=0.
    \label{eq:pole}
\end{equation}
Using Eqs. \ref{eq:u}, \ref{eq:v}, and \ref{eq:pole}, we derive the JJ energy-phase relationship for a transparent barrier,
\begin{equation}
    E=\pm\Delta\sqrt{1-\text{sin}^2\phi/2},
    \label{eq:analDisp}
\end{equation}
plotted in Fig. \ref{fig:SCSC_ABS_numEn} as a dotted line. For the opaque barrier solution, multiply the sine function by $\eta$ which approaches zero for strong barrier potential. Expanding Eq. \ref{fig:SCSC_ABS_numEn} for $\eta\ll 1$ then produces the usual result of $E\propto \pm\text{cos}(\phi)$.

From this analysis, we derive the expected result that out-of-phase SC systems produce two junction bound states. These appear as poles in the $A_{ph,hp}$ coefficients. One interprets this to mean that there is always non zero probability that a p-like wave transmits as a h-like and vice-versa. This is exactly the standard interpretation for the ABS where the bound state exists due to continual exchange of particle and hole waves brought on by Andreev reflection within the junction.

\subsection{Tight Binding Model}
\label{sect:scscTB}
The previous calculation is a sufficient introduction for the SC-SC junction, but it is cumbersome to apply for systems that are slightly more complicated. For the more general setting, we fall back on a tight-binding approach, but we will demonstrate that this numerical scheme recovers the correct dispersion relation. This method will be used to calculate $E(\phi)$ for the rest of the non trivial junctions with a more complex system of nanowires. Now, we recast our problem in the bottom of Fig. \ref{fig:scscSetup} as a finite system made of $N$ lattice sites with two orbitals each spaced apart by lattice constant $a$ with the model BdeG Hamiltonian,
\begin{align}
    H= \sum^N_n\{-\mu\tau_z^{(n)}\ket{n}\bra{n}-t&\tau_z^{(n)}\ket{n+1}\bra{n}\nonumber\\&+\hat{\Delta}^{(n)} \ket{n}\bra{n}\}+h.c.,\label{eq:SCSCTB}
\end{align}
where $\tau^{(n)}_i$ are $2\cross 2$ Nambu matrices acting on the orbitals at site $n$, $t$ is the hopping parameter, $\mu$ is the onsite chemical potential, and $\hat{\Delta}$ is the onsite pairing operator,
\begin{gather}
    \hat{\Delta}^{(n)} = \Delta_0\begin{pmatrix}
        0&e^{i\phi_n}\\
        e^{-i\phi_n}&0
    \end{pmatrix}
    \propto \tau^{(n)}_x \text{cos}(\phi_n)-\tau^{(n)}_y \text{sin}(\phi_n)\label{eq:pairingOp},
\end{gather}
The onsite phase, $\phi_n$, is zero for all $n<N/2$ and constant for all $n>N/2$. 

Based on the previous section, when solving $H\psi_j=E_j\psi_j$ for the spectrum and states, we anticipate energy levels that appear in the gap of the comparable bulk system. In Fig. \ref{fig:SCSC_specComp}, we verify this claim by plotting the composite finite ($N=30$) system's eigenvalues alongside the bulk spectrum for left and right systems for $\phi = \pi$, $t=\Delta_0=1$ eV, and $\mu=0.5$ eV. One finds that indeed there are two levels which seperate from the bulk spectrum into the gap. The focus of our method is to vary $\phi$ and solve for the system's eigenvalues, keeping track of these in-gap energy levels. The content of our calculation is plotted alongside the analytical result in Fig. \ref{fig:SCSC_ABS_numEn}, where we see that these phase sensitive states oscillate in a sinusoidal way. The gap that emerges here is explained by the fact that a finite sized hopping parameter exists in between the two systems. If we were to increase the junction's $t\rightarrow\infty$, the gap between the two bound states would approach zero, and the calculation would perfectly reproduce Eq. \ref{eq:analDisp}.

To make further commentary on our systems, we introduce a collection of observables to analyze the associated in-gap eigenstates. To this end, let an arbitrary state be decomposed into a sum over site orbitals, $\ket{\psi}=\bigoplus_n \ket{\varphi^n}=\bigoplus_n (\alpha^p_n,\alpha^h_n)^T$, and define the following operators,
\begin{equation*}
    T_z = \bigoplus_n \tau^{(n)}_z,\hspace{10pt}
    T_x = \bigoplus_n \tau^{(n)}_x,\hspace{10pt}
    T_y = \bigoplus_n \tau^{(n)}_y.
\end{equation*}
\begin{figure}[t!]
    \centering
    \includegraphics[width=\columnwidth]{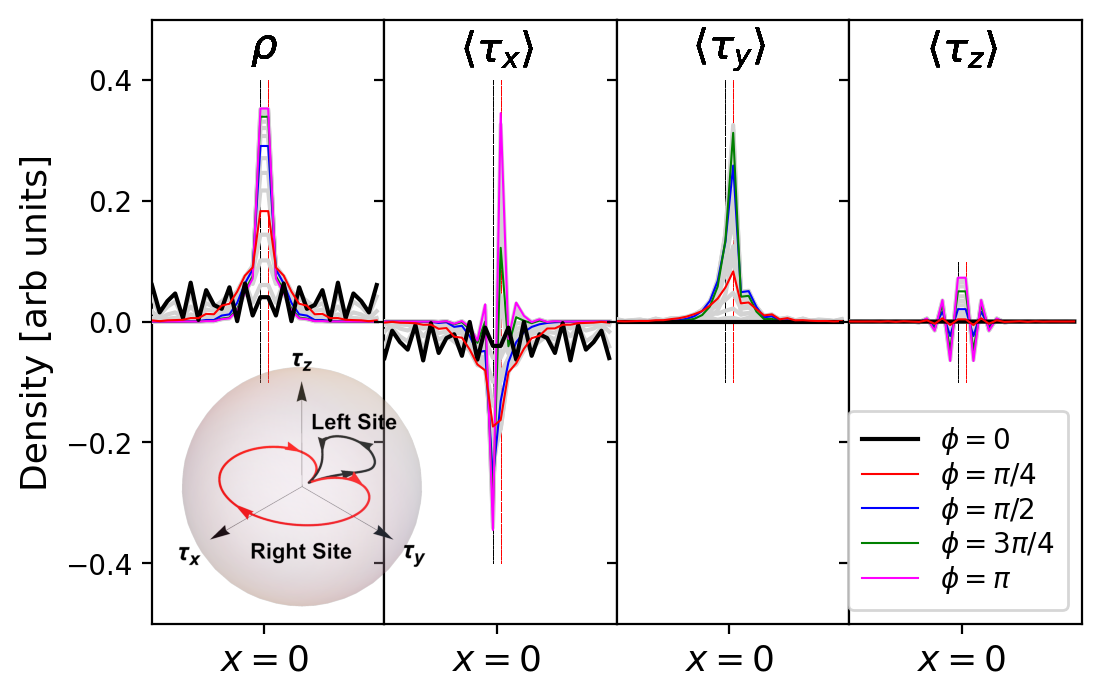}
    \caption{SC-SC Local Densities. The local densities introduced in the text are calculated and plotted here for the lower energy ABS. We select five representative values of $\phi$ to color, and in between values are plotted in gray. The inset shows the trajectory of these quantities for the sites left (black) and right (red) of the junction.}
    \label{fig:SCSC_ABS_dense} 
\end{figure}
Firstly, taking the expectation value of $T_z$, we find that $\bra{\psi} T_z\ket{\psi}=\sum_n\bra{\varphi^n}\tau^{(n)}_z\ket{\varphi^n}=\sum_n (\abs{\alpha^p_n}^2-\abs{\alpha^h_n}^2)$. Thus, $\langle T_z \rangle$ is the total sum of the sitewise difference between particle and hole weights, producing the total charge for a given eigenstate. Thus, we can interpret $\tau^{(n)}_z$ as the \textit{local charge density} for each state. In other words, $\tau^{(n)}_z$ measures the likelihood that a given site is a particle or hole. For example, an arbitrary site with orbital mixture $\ket{\varphi^n} = (1,0)_n^T$ produces $\bra{\varphi^n}\tau^{(n)}_z\ket{\varphi^n}=1$, meaning that site is occupied by a particle orbital only.

From Eq. \ref{eq:pairingOp}, we can immediately see that $T_x$ and $T_y$ are related to the total pairing operator, and thus $T_{x,y}$ measure \textit{global} aspects related to pair production, $\tau^{(n)}_x$ and $\tau^{(n)}_y$ measure \textit{local} pairing qualities. As a concrete example, if a site has orbital mixture $\ket{\varphi^n}=\frac{1}{\sqrt{2}}(1,1)_n^T$ then $\bra{\varphi^n}\tau^{(n)}_x\ket{\varphi^n}=1$, so $\tau^{(n)}_x$ measures the local equality of the absolute value of particle/ hole orbital weights. Thus, we call $\langle\tau^{(n)}_x\rangle$ the \textit{real local quasiparticle density}. Similarly, $\langle \tau^{(n)}_y\rangle$ is the \textit{complex local quasiparticle density}, measuring the site wise phase difference between the orbitals. Finally, in addition to these quantities, we are naturally also interested in the standard probability density $\rho_n = \braket{\varphi^{(n)}}{\varphi^{(n)}}$ which sums to unity $\braket{\psi}{\psi}=\sum_n \rho_n=1$.

In Fig. \ref{fig:SCSC_ABS_dense}, we calculate and plot these quantities with respect to variations in $\phi$ for the low energy ABS. On the far left, we demonstrate the appearance of a bound state localized to the junction for $\phi>0$ with a maximal localization ocurring at $\phi = \pi$, justifying the title of bound state. Coinciding with this behavior, $\langle \tau^{(n)}_x\rangle$ maximally localizes for $\phi=\pi$ as well. This quantity signals that there is a high density of equal magnitude orbital weights around the junction; however, it oscillates negative to positive from left to right. This switching character means that there is first an $(\alpha,\alpha)^T$ mixture then $(\alpha,-\alpha)^T$  following the junction for some real number $\alpha$. In other words, there are junction bound quasiparticles on the edges of both superconductors, one for the left and right system. 

This ABS can be explained in the following way. In the absense of a coupling link between systems, equal energy eigensolutions will possess oscillatory character with the right system's state being out of phase. The result of coupling the two SCs is to cancel the out of phase wavelike $\langle\tau_x\rangle$ everywhere except where a discontinuity exists at the junction. Saying this differently, the two waves cannot coexist simultaneously, so coupling results in zero everywhere except for $x=0$. In fact, these bound states are the tight-binding analogue to those in the continuum model where Andreev reflection generates a continual exchange of particles and holes. In that context, an incident particle wave reflects off of the out-of-phase SC's interface, imparting $\phi$ phase to the reflected hole. Likewise, an electron incident on the left SC gains no phase when reflected as a hole. When this phase difference is anything other than $0$ or $\pi$, we see a pile up of the $\langle \tau_y \rangle$ density at the junction. While processing $\phi$, this quantity becomes maximally positive/ negative for $\phi \cong 3\pi/4$ and $5\pi/4$ respectively. Finally, $\langle \tau_z\rangle$ becomes positively maximal at  the junction, meaning the system develops a \textit{negative} charge at the interface of the two SCs. It should be noted though that the emergence of this charge pileup only exists if $\mu\neq0$. Otherwise, this quantity would be uniformly zero. 

To summarize, the interesting behavior for these states is highly localized to the sites adjacent to the junction, so, we combine the three densities into one image by plotting their values for the left and right of junction within the inset plot of Fig. \ref{fig:SCSC_ABS_dense} over $\phi\in [0,2\pi]$. 

\section{Topological Junctions:}
\label{sect:topoJuncs}
For the essential part of our work, we present here a catalogue of topological nanowire junctions where we augment the previous normal SC junction to include TSC nano wires within the construction. As a note on semantics, we call these systems topological to mean that they are capable of \textit{becoming} topological even if the specific choice in parameters places the system into the trivial regime. We reserve the words \textit{trivial} and \textit{nontrivial} to refer to true topological phase materials.

\subsection{SC-TSC:}
\label{sect:SCTSC}
We adjust the model in Sect. \ref{sect:scscTB} by converting half the system into one capable of becoming topological. For sites $n<N/2$, the model is the same as before; however, for sites $n>N/2$, we swap in Kitaev's chain. The total Hamiltonian decomposes as $H=H_{SC}+H_{TSC}+H_{T}$ with,
\begin{align}
    &H_{SC}= \sum^{N/2}_n \{-\mu\tau^{(n)}_z\ket{n}\bra{n}-t\tau^{(n)}_z\ket{n+1}\bra{n}\nonumber\\
    &\hspace{130pt}+\hat{\Delta}^{(n)}\ket{n}\bra{n}\}\scalebox{0.9}{+h.c.},\nonumber\\
    &H_{TSC}= 
    \sum^{N}_{n>N/2} \{-\mu\tau^{(n)}_z\ket{n}\bra{n}-t \tau^{(n)}_z\ket{n+1}\bra{n}\nonumber\\
    &\hspace{90pt}-i\Delta_0 \tau^{(n)}_y\ket{n+1}\bra{n}\}\scalebox{0.9}{+h.c.},\label{eq:kitChain}\\
    &H_{T}= -V_c\ket{L} \bra{R}\scalebox{0.9}{+h.c.},\nonumber
\end{align}
where $V_c$ couples the two regions at left and right sites at the junction. The important changes are that we allow phase to be variable in the lefthand normal SC region, and the TSC region possesses a hopping order parameter instead of onsite.
\begin{figure}[t!]
    \centering
    \includegraphics[width=\columnwidth]{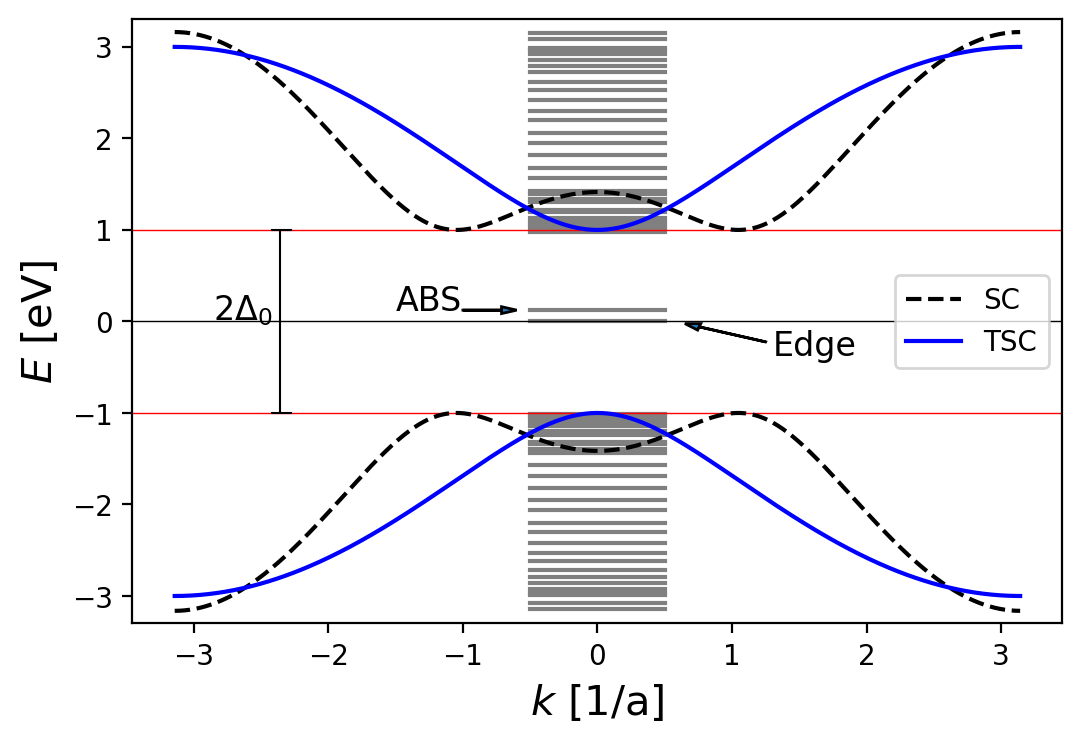}
    \caption{SC-TSC Bulk Dispersion vs. Finite Spectrum. Comparing the bulk dispersion for left and right systems with the finite system spectrum demonstrates there are two in-gap states. There is only one phase sensitive state with the other being locked to zero energy}
    \label{fig:SC_TSCspecComp}
\end{figure}

\begin{figure}[b!]
    \centering
    \includegraphics[width=\columnwidth]{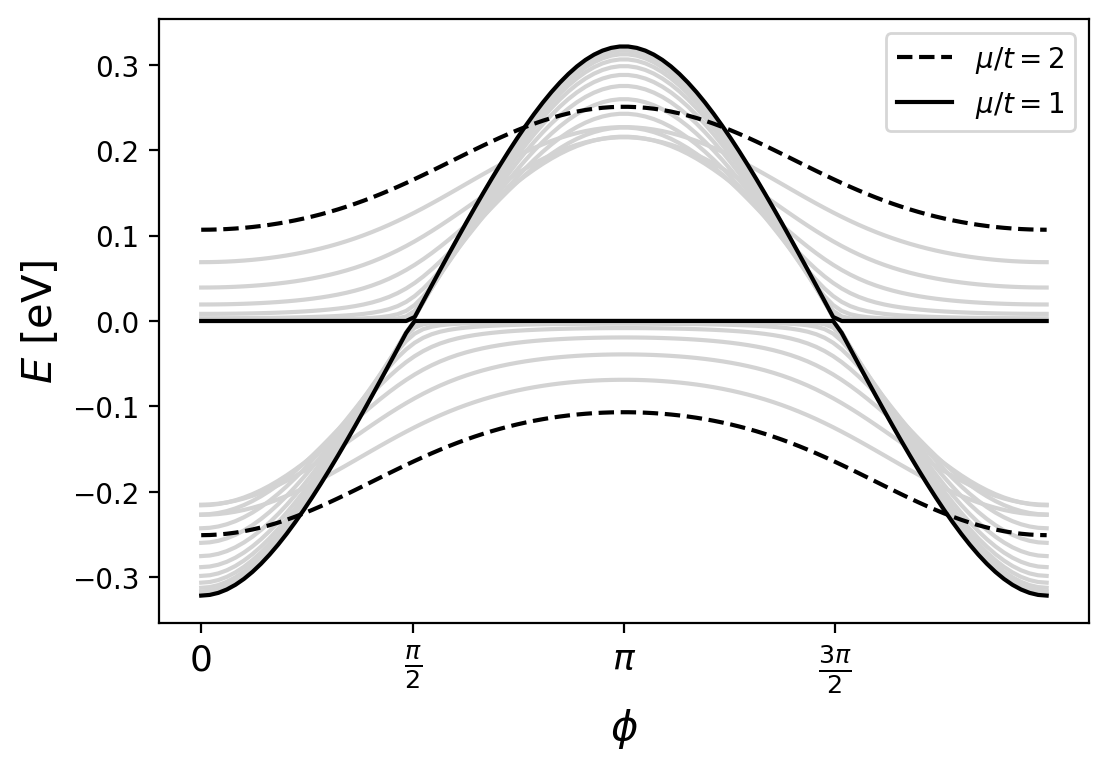}
    \caption{SC-TSC Numerical Calculation of $E(\phi)$. The in-gap dispersion relation is plotted while varying the system parameters. These plots begin in the trivial phase (dotted) and progress into the topological phase (solid). The zero energy state is entirely localized to the boundary of the TSC furthest from the junction, and the phase sensitive state is bound to the junction. Strikingly, we have only one ABS at the junction with zero energy crossings dictated by symmetry}
    \label{fig:SC_TSC_jjEn_tDep}
\end{figure}

Proceeding similarly to the SC-SC analysis, we solve for the bulk spectrum for each region seperately and the finite energy levels for the composite system to retrieve the in-gap eigenstates. This computation is plotted in Fig. \ref{fig:SC_TSCspecComp} for $N=30$ and $t=\mu=\Delta_0=V_c=1$ eV, demonstrating the emergence of two bound states in the gap. In Fig. \ref{fig:SC_TSC_jjEn_tDep}, we extract these two states and iterate their energy over the phase variable for a topologically trivial (dashed) and nontrivial (solid black) system, $\mu/t=2,1$  respectively. The solid gray plots are dispersion curves for systems that lie in between the two extremes. For the topologically trivial system, we find that $E(\phi)$ is different from the SC-SC junction with the main difference being that the upper state reaches maximum and minimum at the same values of $\phi$ as the lower curve. This means that the states are out of sync with one another, where, for $\phi = 0,\pi$, there is always at least one in-gap state. This is different from the in sync SC-SC ABS pair. As the parameters are varied towards a topological phase, there emerges the characteristic zero energy state; however, one finds that there is only a \textit{single} state like this for this system. This zero energy state is also entirely localized to the edge of the TSC uninvolved with the SC-TSC interface. This is different behavior from the standard TSC where there are two zero energy states made of two MBSs each. The phase sensitive bound state is, as expected, localized to the junction. However, it is intriguing to question where the other topological edge state has gone. 

\begin{figure}[t!]
    \centering
    \includegraphics[width=\columnwidth]{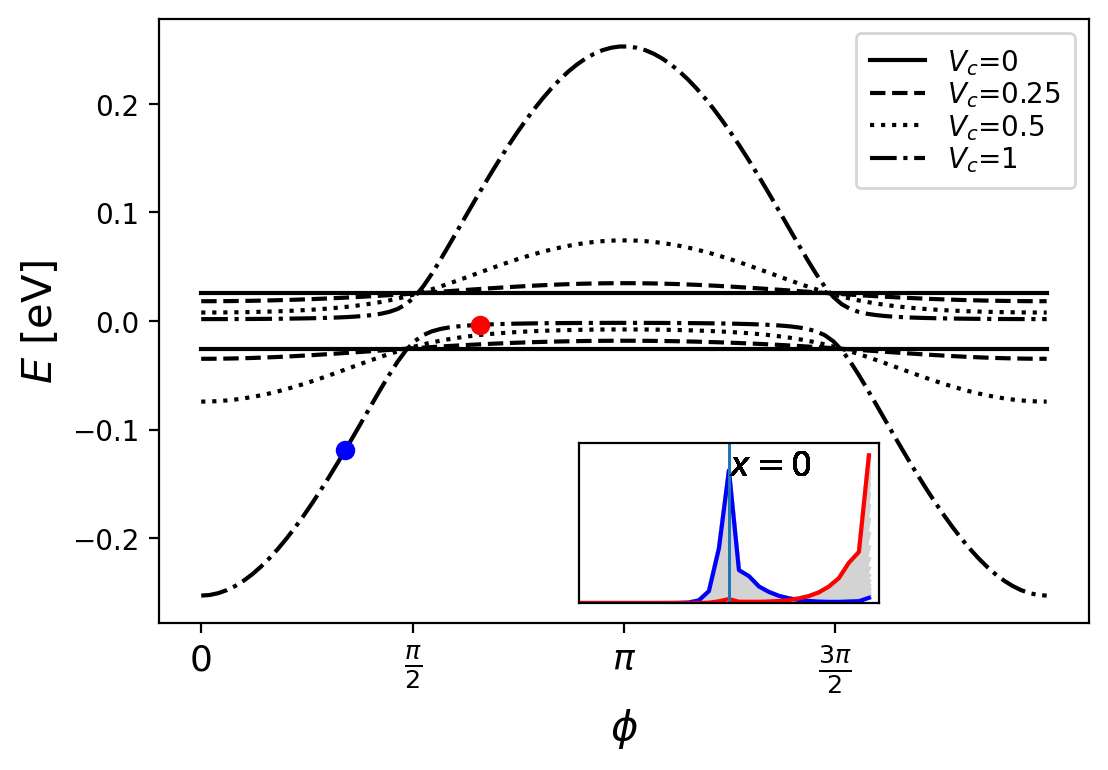}
    \caption{Almost Topological SC-TSC Dispersion. We note that the coupling parameter is responsible for the junction bound state becoming phase sensitive. For the near topological regime, $\mu/t\cong1.429$, we observe a state that switches localization based on phase.}
    \label{fig:SC_TSC_jjEn_Vcdep}
\end{figure}

To probe this question, we place the system in the \textit{almost} topological phase, $\mu/t\cong1.429$, and vary $V_c$ from 0 to 1 eV in Fig. \ref{fig:SC_TSC_jjEn_Vcdep}. We do this to slightly split the degeneracy between the otherwise zero energy states for the TSC. For a maximally decoupled system, $V_c=0$ eV (solid), we recover two, in-gap states with flat dispersion curves which will approach zero for a nontrivial system. For $V_c>0$ eV though, we can see that only certain portions of these curves develop phase dependence. For a fully topological TSC, it is the edge state localized to the junction that becomes the phase dependent state. Thus, we observe that one of the MBS for a TSC is converted into an ABS due to the out-of-phase, adjacent SC. Furthermore, we highlight here an interesting feature of this system while fully coupled and in the \textit{almost} topological regime. In Fig. \ref{fig:SC_TSC_jjEn_Vcdep}, we find that the two bound states develop a gap at $\pi/2$; however, the state retains the expected spatial character for the flat or bell portion of the curve. That is, on the flattened portions of this plot, the state is entirely localized to the non junction edge of the TSC, but, on the curved portions, the state is localized entirely at the junction edge of the TSC. Referencing the inset of Fig. \ref{fig:SC_TSC_jjEn_Vcdep}, as one varies $\phi$ from the blue point to the red point, the state will shift from a junction bound state to the edge bound state. One can imagine that a time dependent $\phi(t)$ will have the affect of oscillating a state back and forth from each location, independent of system size. 

\begin{figure}[b!]
    \centering
    \includegraphics[width=\columnwidth]{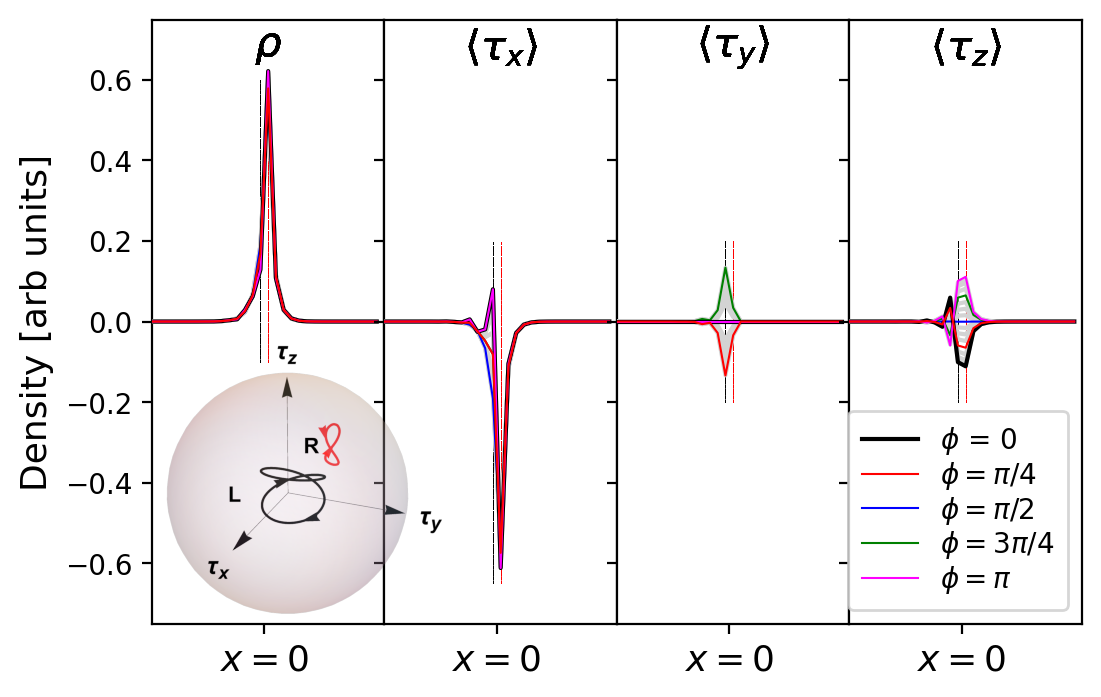}
    \caption{SC-TSC Local Densities. There is a highly localized state at the junction for all $\phi$. Most of the interesting behavior occurs for the junction left site within the SC. }
    \label{fig:SC_TSC_dense}
\end{figure}

Moving on to the local density calculations, we place the system back into the nontrivial phase with parameter combination found in Fig. \ref{fig:SC_TSC_jjEn_tDep} and plot our findings for the single phase sensitive state in Fig. \ref{fig:SC_TSC_dense}. We do not show the zero energy state, because the state's spatial character is invariant with respect to $\phi$. Contrary to the SC-SC system, the junction bound state is strongly localized \textit{for all} values of $\phi$ with most of the $\tau_{x,y}$ variation ocurring at the site immediately to the left of the junction in the normal SC. The $\langle \tau_x \rangle$ density shows very similar slight spatial oscillation behavior left of junction as in Sect. \ref{sect:scscTB}; however, this occurs even for $\phi = 0$ because of the topological barrier provided by the TSC. This state also oscillates in between positive and negative values with phase at the left of junction site, but the right side exhibits the expected behavior for the edges of a TSC, remaining negative throughout $\phi$ variation. Another subtle difference here is that the $\langle \tau_y \rangle$ density reaches both a maximum and minimum for $\phi = \pi/4$, $3\pi/4$ respectively before $\pi$. Likewise, $\langle \tau_z \rangle$ can be seen to oscillate between positive and negative for the single state, meaning the ABS here has access to both positively and negatively charged character along the dispersion curve.

To comment on topological nontriviality, on its own the TSC is nontrivial for all values of $\phi$. This is easily verified by checking the particle-hole antisymmetry ($\mathcal{P}$) of $H_{TSC}$. Here, $\mathcal{P}=T_x\mathcal{K}$ where $\mathcal{K}$ is conjugation, and $\mathcal{P}$-antisymmetry is respected if $\mathcal{P}H\mathcal{P}^{-1}=-H$. This amounts to applying $\tau^{(n)}_x$ to each block of $H$. Since $\tau_x\tau_z\tau_x=-\tau_z$, the regular onsite and hopping terms for both systems obey antisymmetry, and the TSC pairing term, $-i\tau_y$, famously satisfies antisymmetry as well. Thus, we only need to check the regular SC term which transforms like so,
\begin{gather}
    \tau_x\mathcal{K}\begin{pmatrix}
        0&e^{i\phi}\\
        e^{-i\phi}&0
    \end{pmatrix}\tau_x=\begin{pmatrix}
        0&e^{i\phi}\\
        e^{-i\phi}&0
    \end{pmatrix}.
\end{gather}
This means that regular SC pairing is the only \textit{symmetric} term within $H$, and therefore $H$ is not $\mathcal{P}$-(anti)symmetric. However, the interplay between these two systems with their contrasting symmetry constraints no doubt plays a role for the behavior of the resulting ABS and dispersion curves. For example, the dispersion crossing points at $\phi = \pi/2$, $3\pi/2$ correspond to places where the junction bound state and the TSC edge state are degenerate. Referencing Fig. \ref{fig:SC_TSC_dense}, the ABS $\tau_x$ density at these points becomes fully negative and almost symmetric across the junction. These forced crossings coupled with the fact that there is only one ABS present mean that the Josephson current is chiral at these special values of $\phi$.

The density behavior is again summarized within the inset of Fig. \ref{fig:SC_TSC_dense}, where oscillations in $\tau_{x,y,z}$ for both left and right sites form a bow tie orbit. The left site wraps around the origin to reflect the fact that the phase procession is occuring within the SC system, and the right site is nearly constant for $\tau_x$ with most of the motion taking place in the $\tau_{y,z}$ plane. The crossing point for both orbits occur for when the dispersion curve crosses zero.

\subsection{TSC-TSC:}
\label{sect:TSCTSCjunction}
The next logical way to increase complexity is to convert both regions into a TSC by making two copies of Eq. \ref{eq:kitChain},
\begin{align*}
    &H_{TSC}= \sum^{N}_n \{-
    \mu_n \tau^{(n)}_z\ket{n}\bra{n}-t\tau^{(n)}_z \ket{n+1}\bra{n}\nonumber\\
    &\hspace{100pt}-\hat{\Delta}^{(n)}\tau^{(n)}_z \ket{n+1} \bra{n}\}+h.c.,
\end{align*}
where $\hat{\Delta}^{(n)}\tau^{(n)}_z$ is introduced to ensure particle-hole anti symmetry is maintained for all $\phi$.
\begin{figure}[t!]
    \centering
    \includegraphics[width=\columnwidth]{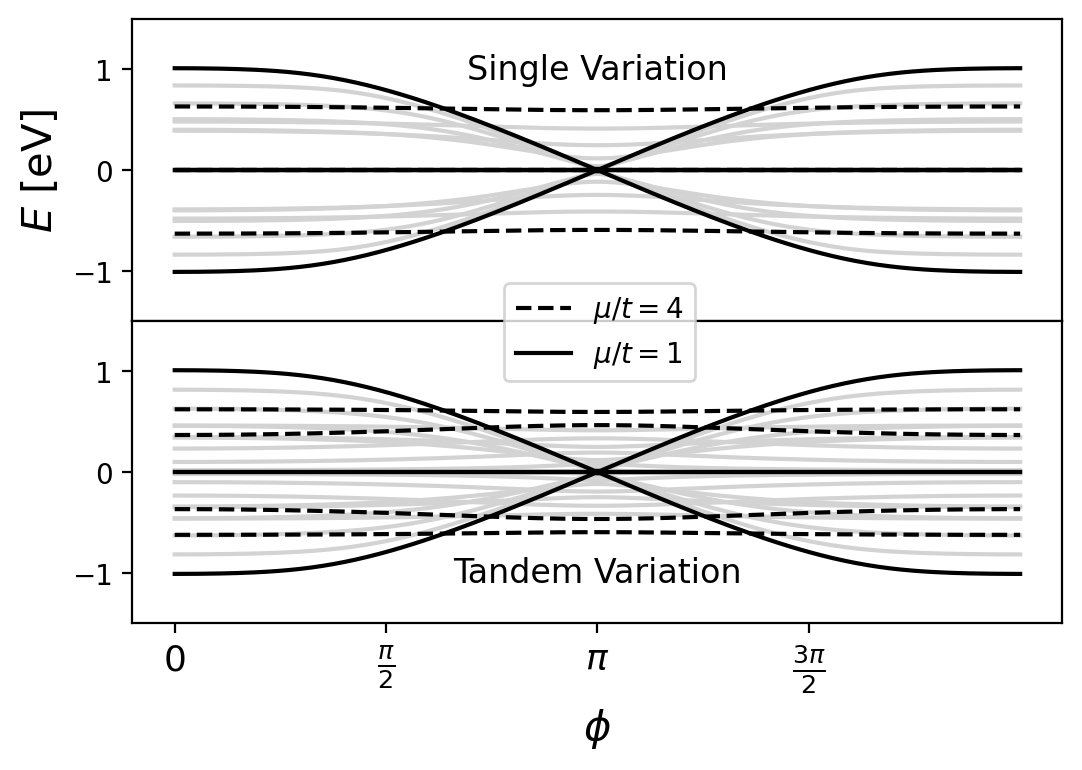}
\caption[TSC-TSC Junction]{TSC-TSC Numerical Calculation of $E(\phi)$. (Upper) Placing one of the systems into the nontrivial regime and varying $\mu/t$, there are four in-gap states with two remaining zero. The other two begin as straight lines which develop into the characteristic sinusoidal curves. (Lower) Varying the topological phase of both systems in tandem shows all four in gap states which then approach either zero or the ABS curves.}
\label{fig:TSC_TSC_jjEnTdiff}
\end{figure}
Setting $\phi_n=0$ for all $n<N/2$, i.e. phase dependence is given to the right TSC, one can see that the TSC model from Sect. \ref{sect:SCTSC} is reproduced. For two TSCs, we can explore a larger parameter space where one or both of the systems can be non trivial, so we also allow $\mu_n$ to be constant within each system but different for left and right regions. For two uncoupled nontrivial systems, each TSC possesses two zero energy, edge localized states. From our analysis in Sect \ref{sect:SCTSC}, we expect that increasing the coupling between systems will have the effect of converting the MBSs located at the junction to ABSs. So, for our energy calculation, we anticipate two phase dependent states at the junction and two zero energy states at the far edges of the composite system. This expectation is validated in Fig. \ref{fig:TSC_TSC_jjEnTdiff} where we have plotted in-gap states as a function of $\phi$ with variations in topological parameters so that one or both of the systems are placed into the nontrivial phase. For this calculation, we have used the same parameter combination as in Sect. \ref{sect:SCTSC} where we vary $\mu/t$ from 4 to 1. This portion of the numerical result is reported in Ref. \cite{prada2020andreev}. Here, we compute the dispersion for the setting where one TSC is always nontrivial (upper) and for the setting where both are varying in tandem (lower) from trivial to nontrivial. For the upper plot, it appears that there are only two in-gap states (dotted) that become ABS; however, there are actually two more dotted lines that remain at zero, i.e. there are always zero energy states native to the nontrivial phase TSC. For both the upper and lower plots, we note behavior that has not been exhibited in the previous sections though. The dotted curves are relatively flat here, meaning these states are mostly invariant with respect to changing $\phi$, and the two phase sensitive states only become aware of $\phi$ once one or both of the TSCs enters the topological regime, see black curves. This behavior is similar to the observation that coupling must become nonzero for there to be an ABS, but here the situation is slightly different. For both of our calculations here, the systems are totally aware of each other, but it is only when nontrivial topology is present that the ABS are as well. This could be due to the fact that the phase variation is handled by a hopping term rather than an onsite term as in the previous two systems. 
\begin{figure}[bb!]
    \centering
    \includegraphics[width=\columnwidth]{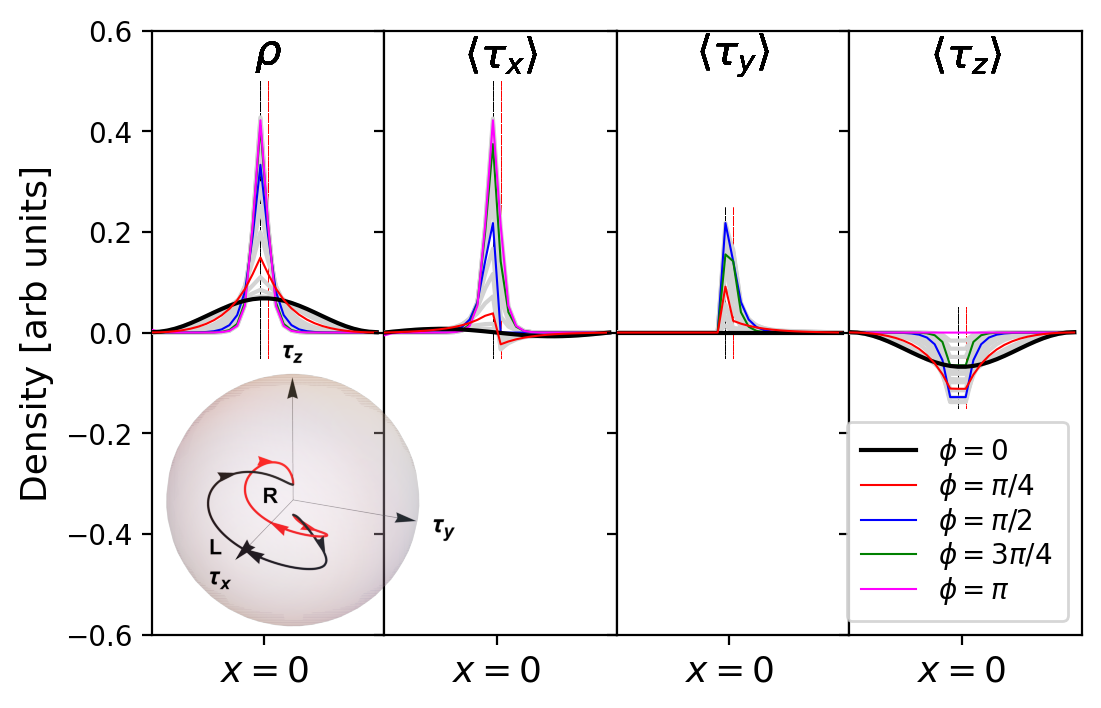}
\caption[SC-TSC Junction]{TSC-TSC Local Densities. Due to the crossing points, we see here that both ABS states are forced to become one another upon $2\pi$ procession of $\phi$. Therefore, the densities depicted here and the inset orbits do not end where they began.}
\label{fig:TSC_TSC_dense}
\end{figure}

Similar to Sect. \ref{sect:SCTSC}, the dispersion crossing point constrains the behavior of the densities because the two curves must become one another after processing $\phi$ through $2\pi$. In Fig. \ref{fig:TSC_TSC_dense}, we plot the characteristic densities for the ABS alongside the combined trajectories for the left and right of junction sites in the inset for the ratio $\mu/t=1$. Both $\langle \tau_{x,y}\rangle$ ABS densities are identical for all values of $\phi$, but the curve's $\langle \tau_z \rangle$ are negative of one another, meaning, whenever the energy curve is positive/ negative, there is a pileup of negative/ positive charge at the junction. Interestingly, because of the gap closure, the charge must oscillate from positive to negative for both states. For the special gap closure point, we can also see here that $\tau_x$ appears as though there is an edge state for each TSC, meeting at the junction almost as if the system is behaving like two disconnected TSC nanowires. 

\subsection{MSQ:}
\begin{figure}[t!]
    \centering
    \includegraphics[width=\columnwidth]{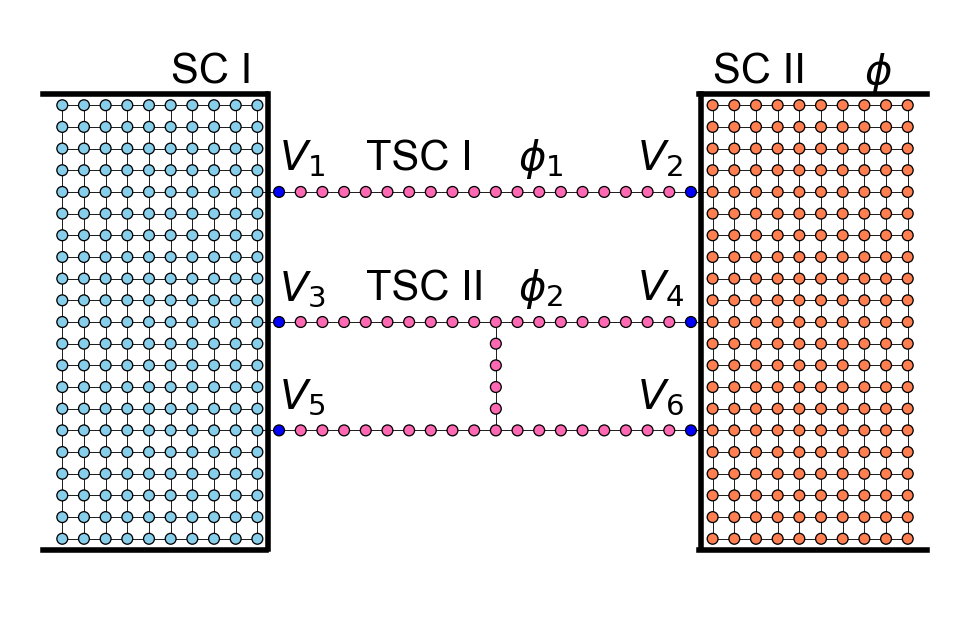}
\caption[SC-TSC-SC Junction.]{MSQ System. The MSQ is comprised of two TSC nanowire islands sitting in between a standard JJ. The two TSC islands have their own phase values, $\phi_{1,2}$, and there are six coupling parameters, $V_i$. We choose to set $V_3=V_6=0$ to model an experimental setting which constrains the allowed current that passes through the structure.}
\label{fig:MSQsetup}
\end{figure}
For the final junction, we model a specialized system proposed by Fu and Kane 2018 in Ref. \cite{schrade2018majorana} called the Majorana SC qubit (MSQ) junction, see Fig. \ref{fig:MSQsetup} where the junction is 2D with two TSC islands, one wire and one ``I" shaped, in between two normal SC systems. Using \texttt{Kwant} \cite{groth2014kwant}, we build a model with 20$\cross$30 sites for two host SCs connected at six locations by two TSC nanowire structures which are each 20 sites long. These six coupling points are associated with voltage gates named $V_{i}$. For our calculation, we set $V_{3,6}=0$ eV and all others equal to 1 eV to emulate a device configuration where current would be forced to travel through point contacts 4,5. We also set $\mu_{SC}=\mu_{TSC}=0.25$ eV, $t=0.5$ eV, and $\Delta_0=1$ eV for the remainder of this section. There are now three phase parameters, $\phi_i$, one for each TSC as well as the right most SC with the left SC phase being zero again. This construction is a highly attractive option for the development of fault tolerant qubits because, despite its nontrivial shape, it really is a modification of a normal JJ with TSC nanowire architecture. Despite this, it has not gained much attention presumably because its nontrivial shape makes rigorous qubit dynamics and error analysis difficult to accomplish. Within the original proposed work, the authors put forth a convincing proof of concept using a tunneling matrix and perturbative type argument. While perfect for an initial, top level assessment, this method lacks the ability to analyze the energy-phase relation, one of its drawbacks \cite{martinis2004superconducting,titov2006josephson}. For this reason, we apply our computation to the full system to numerically derive the structure of $E(\phi)$.
\begin{figure}[b!]
    \centering
    \includegraphics[width=\columnwidth]{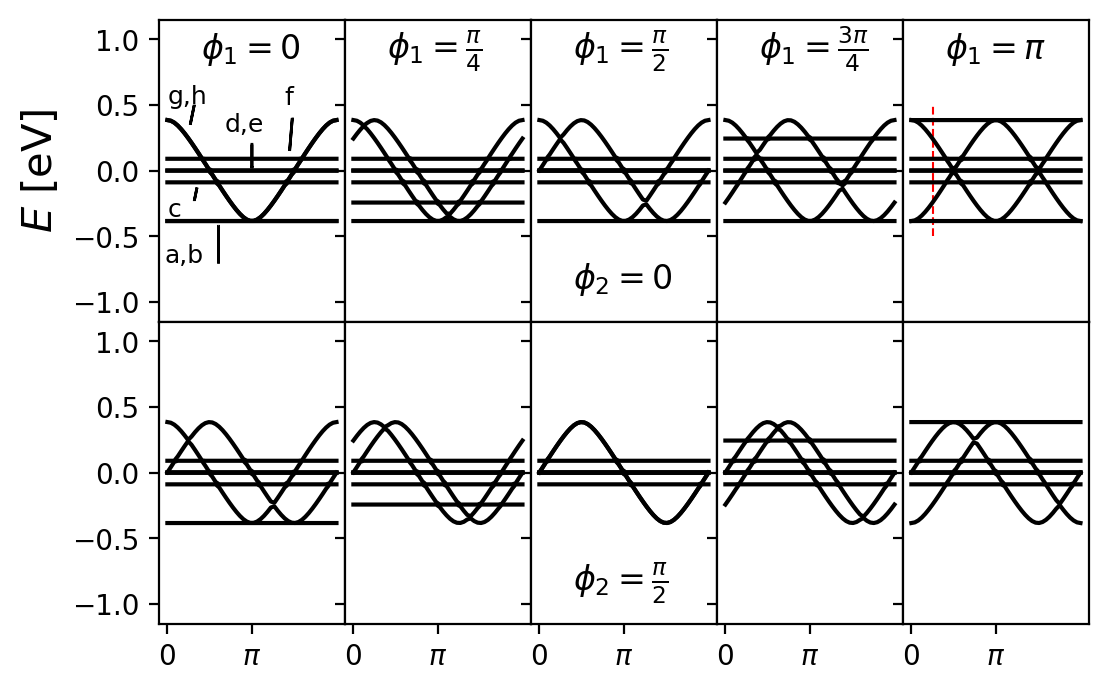}
    \caption{MSQ Numerical Calculation of $E(\phi)$. For the $V_3=V_6=0$ configuration, we see two $\phi$ sensitive states (g,h) for sites nearest to the right SC. There are two bridge states (c,f) symmetric about zero, and two near zero energy states (d,e) resulting from the zero couplings. There are also two flat nonzero curves (a,b) that are $\phi_{1,2}$ sensitive respectively.}
    \label{fig:MSQJJEn}
\end{figure}

\begin{figure}[t!]
    \centering
    \includegraphics[width=\columnwidth]{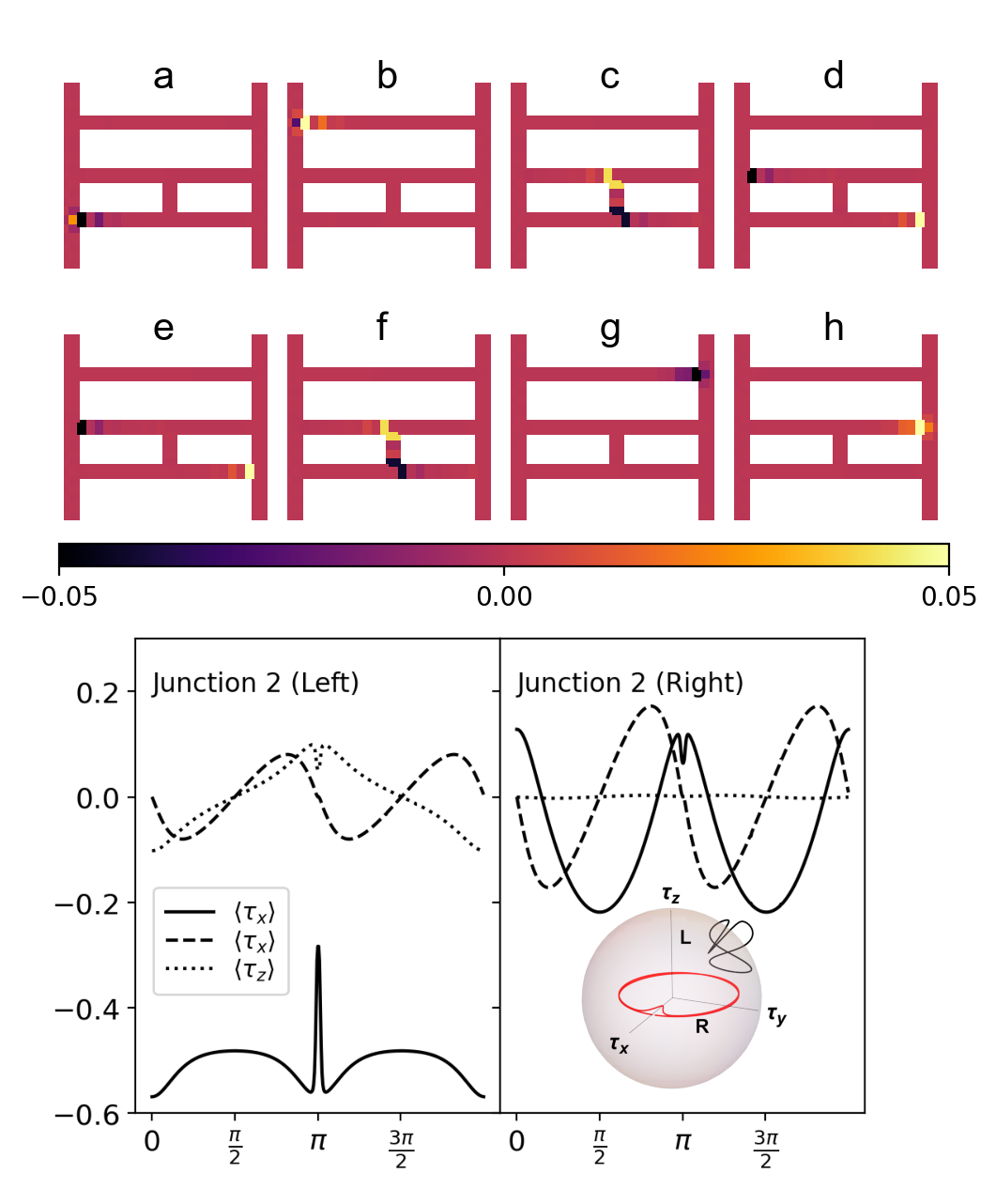}
    \caption{MSQ Local Densities. (Upper) The $\langle \tau_x\rangle$ is plotted for $\phi=\pi/4$, $\phi_1 = \pi$, and $\phi_2=0$ to depict where they are located. (Lower) We plot the trajectories of the densities for sites nearest to coupling $V_2$ with $\phi_1 = 0$.}
    \label{fig:MSQJJDense}
\end{figure}

A system such as this will have several in-gap bound states compared to the 2D SC bulk values. However, many of these correspond to the TSC nanowire's presence as a finite system and are not interesting here. In Fig. \ref{fig:MSQJJEn}, we plot the energy-phase relationship for the eight interesting bound states while varying $\phi_1$ for each column and $\phi_2=0$, $\pi/2$ for the rows. Beginning with the bottom curve of the top left plot where $\phi_{1,2}=0$, there are two overlapping flat states (a,b) which correspond to state localization to sites 1 and 5, the TSC edges that make contact with the left SC. Next, there are four more flat curves that are symmetric about zero. The two furthest from zero (c,f)  both correspond to the added bound states that occur when a TSC nanowire meets another at some mid point. This is a known fact of TSC nanowire structures that whenever one TSC is attached at a mid point of another, there emerges bound states similar to the normal edge states \cite{tutschku2020majorana,stanescu2018building,huang2018quasiparticle}. The other two flat curves closest to zero (d,e), but not actually zero with $E = \pm 1.31\cross 10^{-3}$ eV, are edge states localized to sites 3 and 6. These are closest to zero because we have turned these couplings off. Finally, there are two overlapping, phase sensitive ABSs (g,h) that are converted via the same process in previous sections and are located at sites 2 and 4 near the right phasing varying SC. Examples of the $\langle \tau_x\rangle$ density can be found in Fig. \ref{fig:MSQJJDense} with the configuration denoted by the red dotted line in the top right panel of Fig. \ref{fig:MSQJJEn}.

Moving on to the second column of this figure where $\phi_1 = \pi/4$, we see two curves are now shifted from there previous positions. One of the two previously degenerate flat curves moves up slightly and one of the ABS shifts out of phase with the other. Since changes to $\phi_1$ only affect TSC I, we know that these shifting curves correspond to the bound states located at the left and right edges of the TSC wire. As we continue to vary $\phi_1$, the ABS curve shifts to the right as expected from known SQUID dynamics where the standard JJ energy contains a phase offset parameter. The flat curve shifts up reaching a maximum that coincides with the ABS curves where further increases in $\phi_1$ will force this state back down to its original position at $\phi_1=2\pi$. In Sects. \ref{sect:SCTSC} and \ref{sect:TSCTSCjunction}, we noted that there would be special gap closure for the ABS dispersion relationship based on $\phi$. Here too, the crossing points of the ABS with the flat curves represent important values of $\phi$. 

For example, observe the top row for $\phi_1=\pi/2$ (middle panel) where the flat curve associated to state localization at site 1 seemingly disappears. It actually passes directly through $E=0$ eV for all $\phi$. We can see for this situation that the TSC I ABS passes through zero at $0,\pi/2,\pi$ and $3\pi/2$. At $0$ or $\pi$, there are true MBSs located at sites 1 and 2, and, for $\pi/2$ or $3\pi/2$, localized to 1 and 4. In other words, for this specific combination of parameters, one can vary $\phi$ to nucleate MBS in the 1,2 combination on the wire portion of the TSC, but further variation of phase leads to the nucleation for the 1,4 combination. Furthermore, the second row shows that setting $\phi_2=\pi/2$ likewise places the flat curve associated to the site 5 bound state at zero energy, so, for the middle plot, second row where $\phi_{1,2}=\pi/2$, one will find gap closure and MBSs for each edge of both TSCs at $\phi=\pi$.  

Lastly, we depict the procession of the local densities for left and right sites of junction 2 alongside the 3D orbits within the lower image in Fig. \ref{fig:MSQJJDense}. The right side (SC) shows no charge build up ($\tau_z=0$), remaining in the $\tau_{x,y}$ plane for all $\phi$. The orbit is a flattened bow tie shape that is almost perfectly circular, slightly off centered, and there is a feature at $\phi = \pi$ which slightly approaches the origin. This means that the localization of the bound state is nearly constant until the two systems are $\pi$ out of phase. At this value, the state spreads out slightly. The circularity of the orbit is simply a by product of the phase procession occuring for the SC. The left of junction site (TSC I) does not orbit around the origin because we are not changing $\phi_{1}$; however, it does oscillate back and for in the $\tau_y$ direction, forming a similar bow tie shape from the SC-TSC system. Since the TSC is non trivial, there will be a strongly localized bound state for most values of $\phi$ which is why the trajectory remains highly negative in the $\tau_x$ direction, but, again, for $\phi=\pi$, one can see this density approach zero sharply. This coincides with the $g$ and $b$ energy levels meeting one another which are associated to the TSC I right and left edge bound state respectively, signaling the slight delocalization of the right hand ABS. 

\section{Towards Fault Tolerance}
This work is a small part of an effort to improve the current hardware capabilities where we report the ABS dispersion relationship for a collection of nanowire devices. These findings can be used in a full circuit analysis where the nanowire architectures have modified, say, the SQUID within a transmon qubit. The energy contribution of this junction is to be used to not only address the perturbative circuit dynamics, but also to analyze the characteristic times related to qubit performance. Our main findings are the various $E(\phi)$ plots throughout as well as the qualitative behavior of the bound states with respect to the topological phases of the systems.

Qubit hardware has seen numerous evolutions since the inception of QC, and it seems that the SC qubit, the so called artificial atom, is so close to what might be the fault-tolerant qubit of the future. It is no wonder that many within the private sector are pushing forward with compact chip designs even with their faulty nature. So far, the main method of improving coherence times has been due to clever tricks regarding the annealing process or sophisticated error correction schemes. However, these strategies can only push the SC type hardware so far, and we should shift our strategies in order to make great changes. Topological materials are extremely attractive for this very purpose. After all, at every stage of microelectronic development, the inclusion of new material classes has drastically reshaped our capabilities. First, semiconductors launched us into the modern era of solid state electronics, then, superconductors propelled us even further into the NISQ era. Topological materials could very well be the material that leads us out of noisy QC into the fault tolerant era of the future. 
\bibliographystyle{apsrev4-2}
\bibliography{aps}
\clearpage
\appendix

\section{Sturm-Louiville General Method}
\label{app:SLGenMeth}
We lay out here the general method for constructing a Green's  function for a Sturm-Louiville operator which applies to a wide variety of problems,
\begin{equation}
    (p(x)\psi')'+q(x)\psi(x)=f(x),
\end{equation}
on the interval $[a,b]$ with $\psi(a)=\psi(b) = 0$. The Green's  function for any operator is the solution to the inhomogenous problem, 
\begin{equation}
    \mathcal{L}(G(x,x'))=(p(x)G'(x,x'))'+q(x)G(x,x')=\delta(x-x').
\end{equation}

For this type of differential operator under the given BCs, one can construct the solution in the following way,
\begin{equation}
    G(x,x') =  \begin{cases} 
      A\psi_<(x)\psi_>(x') & x< x' \\
      A'\psi_<(x')\psi_>(x) & x'<x 
   \end{cases},
\end{equation}
where $\psi_{\lessgtr}(x)$ are solutions which obey the left ($<$) and right ($>$) BCs respectively. To determine constants $A$ and $A'$, one integrates the differential equation with respect to $x$ from $x'-\epsilon$ to $x'+\epsilon$ then take the limit as $\epsilon\rightarrow 0$. After integrating, the right hand side becomes unity because of the Dirac delta, and the $q(x)$ term goes to zero after the limit is taken. Thus, the total equation to consider becomes,
\begin{align}
    \int_{x'-\epsilon}^{x'+\epsilon} dx (p(x)G'(x,x'))'=1.
\end{align}
From this, one derives the \textit{jump condition},
\begin{equation}
    p(x)[G'(x,x')|_{x'+\epsilon}-G'(x,x')|_{x'-\epsilon}=1.
\end{equation}
Upon subsequent integration, one derives the continuity condition as well,
\begin{equation}
    G(x'+\epsilon,x')=G(x'-\epsilon,x'),
\end{equation}
which sets $A=A'$. The jump condition sets $A=\{p(x)W\}^{-1}$ where,
\begin{equation}
    W=\psi_<(x)\psi'_>(x)-\psi'_<(x)\psi_>(x),
\end{equation}
is the Wronskian.

For an example, consider the operator $\mathcal{L}=-\frac{\hbar^2}{2m}\frac{\partial^2}{\partial x^2}-E$ with open boundary conditions. To solve the inhomogenous problem, $\mathcal{L}(g_0(x,x'))=\delta(x-x')$, one uses the leftward and rightward propogating solutions to the problem, $\psi_\lessgtr(x) = e^{\mp ikx}$. These are obtained by solving the problem on a finite domain and then sending the boundaries to infinity within a limiting procedure. For a helpful mental picture, we think of the $\delta(x-x')$ as a source at point $x'$ within the domain which produces wavelets that travel to the left and right. The mental image of a dense stone dropping into 1D water is extremely helpful here. These wavelets are exactly the left and right solutions to the problem,
\begin{equation}
    g_0(x,x') = A\begin{cases} 
      e^{-ikx}e^{ikx'} & x< x' \\
      e^{-ikx'}e^{ikx} & x'<x 
   \end{cases}.
\end{equation}
We then calculate $A$ by solving for the Wronskian,
\begin{equation}
    -\frac{\hbar^2}{2m}A[(ik)e^{-ikx'}e^{ikx'}-(-ik)^{-ikx'}e^{ikx'}]= 1,
\end{equation}
and sets $A = \frac{i m}{k\hbar^2}$.

\section{Delta Potential}
\label{app:deltaPot}
We will now derive the Green's function for a Dirac potential to demonstrate the method for Sect. \ref{sect:SCSCjunction}. As outlined in Appendix \ref{app:SLGenMeth}, one begins the derivation by constructing states which carry left and right BCs for $\mathcal{L}\Psi = 0$ where $\mathcal{L}=-\frac{\hbar^2}{2m}\frac{\partial^2 }{\partial x^2}+ V_c\delta(x)-E$. One can approach the problem from two equivalent angles. On one hand, because of the discontinuity at $x=0$ brought on by the $\delta$ function and depending on the position of $x'$, our left/right wave solutions will be piece wise with scattering coefficients for incidence, reflection, and transmission. One can solve for these coefficients by imposing conditions on $\Psi$ at $x=0$ and then construct $G(x,x')$. However, since $G(x,x')$ is ultimately formed by the piecewise multiplication of wavelets, one can form a sum out of fundamental wave processes weighted by unknown coefficients. They are set by the jump and continuity conditions on $G(x,x')$. Thus, we can circumvent the scattering coefficient calculation slightly.

To make this point clear, we proceed down the first route now. For both regions, we immediately know that the total state will be linear combinations of wave like solutions, $\psi(x)=e^{\pm i k x}$. Let's say that an incident wave approaches the delta with amplitude, $A$. Clearly, it will scatter off the delta, producing a reflected wave and a trasmitted wave with coefficients $B$ and $C$ respectively. Thus, the total wave function is piecewise,
\begin{equation}
    \Psi(x) = \begin{cases} 
      Ae^{ikx}+Be^{-ikx}& x< 0 \\
      Ce^{ikx} & x>0 
   \end{cases}.
\end{equation}
To fix these coefficients, one integrates $\mathcal{L}\Psi = 0$ around $x=0$, which produces the following conditions at $x=0$,
\begin{align}
    [\frac{\partial}{\partial x}\Psi(0^+)-\frac{\partial}{\partial x}\Psi(0^-)]&=\frac{2mV_c}{\hbar^2}\Psi(0^+)\\
    \Psi(0^+)&=\Psi(0^-).
\end{align}
Using these conditions, one produces the following set of algebraic equations on the coefficients,
\begin{align}
    A-B&=(1+2i\eta)C\\
    C&=A+B,
\end{align}
where $\eta=\frac{mV_c}{k\hbar^2}$. These equations yield the reflection and transmission ratios, $B/A = -\frac{i\eta}{1+i\eta}$ and $C/A = \frac{1}{1+i\eta}$. Notice, these coefficients contain a pole in there analytic continuation; however, in order to identify this pole as a bound state, one must extract the information from the Green's  function itself.

With that, let us now construct the Green's  function to the problem. Since $\Psi$ is piecewise, $g(x,x')$ will come with many pieces depending on the relative location of the source point, $x'$, and the test point, $x$. For example, if $x'<0$, the left BC wave function is simply a wave propogating to the left, $\Psi_<(x)=e^{-ikx}$, but the right BC wave function is the piece wise function from the scattering problem. Thus, for perspective, the total function can be written as,
\small
\begin{equation}
    g(x,x') = \begin{cases}
    A\psi^I_<(x)\psi^{I}_>(x')+B\psi^{I}_<(x)\psi^{I}_<(x')&x<x'<0\\
    A\psi^I_<(x')\psi^I_>(x)+B\psi^I_<(x')\psi^I_<(x)&x'<x<0\\
    C\psi^I_<(x)\psi^{II}_>(x')&x<0<x'\\
    C\psi^{II}_<(x')\psi^{I}_>(x)&x'<0<x\\
    D\psi^{II}_<(x)\psi^{II}_>(x')+F\psi^{II}_>(x)\psi^{II}_>(x')&0<x<x'\\
    D\psi^{II}_<(x')\psi^{II}_>(x)+F\psi^{II}_>(x')\psi^{II}_>(x)&0<x'<x
    \end{cases},
\end{equation}
\normalsize
where $\psi^{I,II}_\lessgtr(x)$ are the basic left (right) waves in region I (II). Since we have already calculated the scattering matrix coefficients $B/A$ and $C/A$, we would have completed our calculation here with the exception of the undetermined $A$. One can clearly see now that the poles in the scattering coefficient do indeed correspond to poles in $g(x,x')$. Crucially, notice also that the structure of $g(x,x')$ is a sum of rudimentary scattering processes weighted by complex coefficients, where $A, D$ corresponds to propogation away from source in same region (incidence), $B, D$ are propogating towards the source (reflection), and $C$ is propogation from left to right of the Dirac potential (transmission). Instead of solving the scattering problem involving $\Psi(x)$ and then constructing $g(x,x')$, we can build $g(x,x')$ up using these basic processes and solve for the coefficients using the jump and continuity conditions. This observation points us toward an alternate route to obtain the coefficients. 

So, let us solve for the coefficients by using the jump and continuity conditions. By integrating the inhomogenous equation $\mathcal{L}(g(x,x'))=\delta(x-x')$, one arrives at the following equation for $x'\neq 0$,
\begin{equation}
    -\frac{\hbar^2}{2m}[\frac{\partial}{\partial x}g(x,x')|_{x'+\epsilon}-\frac{\partial}{\partial x}g(x,x')|_{x'-\epsilon}=1,
\end{equation}
and, for $x'=0$,
\begin{equation}
    -\frac{\hbar^2}{2m}[\frac{\partial}{\partial x}g(x,0)|_{\epsilon}-\frac{\partial}{\partial x}g(x,0)|_{-\epsilon}+ V_cg(0,0)=1.
\end{equation}
The first equation is identical to the one from the free field in Appendix \ref{app:SLGenMeth} which sets only $A=D=\frac{im}{k\hbar^2}$. Since this equation involves the Wronskian, pieces which are travelling in the same direction end up yielding zero here, i.e. the $B$ and $F$ terms which contain functions that are not linearly independent (in fact the functions are identical). The second equation involves the jump at $x'=0$ and will set $C$,
\begin{equation}
    -\frac{\hbar^2}{2m}(2ik))C+ V_cC=1,
\end{equation}
which leads to,
\begin{equation}
    C = \frac{1}{V_c-i\frac{k\hbar^2}{m}},
\end{equation}
and, one can readily confirm,
\begin{equation}
    C/A = \frac{1}{\frac{im}{k\hbar^2}(V_c-i\frac{k\hbar^2}{m})}=\frac{1}{1+\frac{imV_c}{k\hbar^2}},
\end{equation}
the very same ratio from the scattering version of the problem. Then, $B/A$ can be found via continuity of $g(x,x')$ at $x'=0$, $A+B=C$, the same equation from the scattering problem.  

Notice, that the Green's  function calculation immediately returns the correct form of $C$ without the use of an algebraic set of equations. In fact, if we are solely concerned about the analytic structure of $g(x,x')$, then the jump condition at $x'=0$ calculation is all one needs to derive this information! 

\end{document}